# Molecular Design Using Signal Processing and Machine Learning: Time-Frequency-*like* Representation and Forward Design

Alain B. Tchagang, *Member, IEEE*, Ahmed H. Tewfik, *Fellow, IEEE*,
and Julio J. Valdés, *Senior Member, IEEE*

***Abstract*—Accumulation of molecular data obtained from quantum mechanics (QM) theories such as density functional theory ($DFT_{QM}$) make it possible for machine learning (ML) to accelerate the discovery of new molecules, drugs, and materials. Models that combine QM with ML (QM↔ML) have been very effective in delivering the precision of QM at the high speed of ML. In this study, we show that by integrating well-known signal processing (SP) techniques (i.e. short time Fourier transform, continuous wavelet analysis and Wigner-Ville distribution) in the QM↔ML pipeline, we obtain a powerful machinery (QM↔SP↔ML) that can be used for representation, visualization and forward design of molecules. More precisely, in this study, we show that the time-frequency-*like* representation of molecules encodes their structural, geometric, energetic, electronic and thermodynamic properties. This is demonstrated by using the new representation in the forward design loop as input to a deep convolutional neural networks trained on $DFT_{QM}$ calculations, which outputs the properties of the molecules. Tested on the QM9 dataset (composed of 133,855 molecules and 19 properties), the new QM↔SP↔ML model is able to predict the properties of molecules with a mean absolute error (MAE) below acceptable chemical accuracy (i.e. MAE < 1 Kcal/mol for total energies and MAE < 0.1 ev for orbital energies). Furthermore, the new approach performs similarly or better compared to other ML state-of-the-art techniques described in the literature. In all, in this study, we show that the new QM↔SP↔ML model represents a powerful technique for molecular forward design. All the codes and data generated and used in this study are available as supporting materials at the following website: https://github.com/TABeau/QM-SP-ML.***

## I. Introduction

DESIGNING drugs and materials with the properties we dream off is the ultimate goal of many chemical, agrochemical and pharmaceutical industries. Throughout the ages, researchers have come up with different strategies to tackle this challenge. That is, designing molecules with targeted properties. Among these techniques, trial and error approaches which are still used today emerge as the most time consuming and costly process [1]. At the beginning of last century, breakthrough in quantum mechanics (QM) and molecular design (MD) have attempted to solve this problem more scientifically, by solving the Schrodinger equations (SE), which govern the system dynamic at the atomic scale [2]. This equation is very difficult to solve for large systems, and has given rise to the development of a variety of approaches for approximately solving the SEs [2]-[13]. Although these approximate methods are able to reach the chemical accuracy of 1 kcal/mol for total energies and 0.1 ev for orbital energies required for computational MD, they are still very time consuming and calculations may take days depending on the size of the molecules and systems. Ideally, a drug or material designer would like to make quantitative estimates in the chemical compound space (CCS) at reasonable computational cost (i.e. milliseconds per compound or faster) [14]. This is very difficult to achieve using trial and errors or computational QM *ab initio* approaches.

Molecular databases [15]-[18] derived from Density Functional Theory ($DFT_{QM}$) offer new directions, among which new methodologies based on machine learning (ML) [19]-[57]. These techniques known as QM↔ML models have shown great potentials, achieving the same precision as $DFT_{QM}$ at a much lesser computational cost. QM↔ML on its own face different modeling problems, among which the representation of molecules in a way that makes forecast of molecular properties realistic and precise [19]. This question has already been comprehensively addressed in the cheminformatics and

Manuscript received May 4, 2020; revised XXXX XX, 2020; accepted XXX XX, 2020. Date of publication XXXX XX, 2020; date of current version XXXX XX, 2020. The associate editor coordinating the review of this manuscript and approving it for publication was Dr. XXXXXXXX. This work was supported by the National Research Council of Canada, Under the AI4D Program.

A. B. Tchagang is with the Digital Technologies Research Centre, National Research Council of Canada, 1200 Montréal Road, Ottawa, ON, K1A 0R6 Canada (e-mail: alain.tchagang@ nrc-cnrc.gc.ca).

A. H. Tewfik is with the Electrical and Computer Engineering Department, University of Texas at Austin, Austin, TX 78712 USA, (e-mail: tewfik@austin.utexas.edu).

J. J. Valdés is with the Digital Technologies Research Centre, National Research Council of Canada, 1200 Montréal Road, Ottawa, ON, K1A 0R6 Canada (e-mail: julio.valdes@ nrc-cnrc.gc.ca) quantitative structure property relationships (QSPRs) literature, and many molecular descriptors have been suggested [58]. Unluckily, they often require significant amount of domain knowledge and they are not always transferable across the entire CCS [14, 56].

In this paper, we follow the same approach introduced in [19, 20], and adopted by several other authors [14, 57]. We learn the forward mapping between molecules and their energetic, thermodynamic and electronic properties using the Coulomb matrix (CM). The CM is directly derived from the geometry (i.e. structure) representation of molecules and has been shown to be a strong candidate for molecular descriptors. The CM is invariant to translation and rotation but not to permutations or re-indexing of the atoms. Several techniques have been developed in the literature to tackle this concern. Few examples comprise Coulomb sorted Eigen-spectrum [56], Coulomb sorted L2 norm of the matrix's columns [20], Coulomb bag of bonds [23], association of CM with the atomic composition of molecules [53], and random Coulomb matrices [14]. It turns out that some derivatives of the CM such as the Coulomb sorted Eigen-spectrum or Coulomb sorted L2 norm of the matrix's columns is a 1-dimension (1D) order numerical sequence representation of a molecule. From the signal processing (SP) perspective, it can be treated as a 1D signal [59].

Here, we explore a new representation of molecules based on the aforementioned 1D signal (Eigen-spectrum) derived above. The 1D signal is transformed into a time-frequency-*like* (TFL) representation using techniques such as Short Time Fourier Transform (STFT), Continuous Wavelet Transform (CWT) and Wigner-Ville distribution (WVD). We show that these 2D TFL representation of molecules encode their structural, geometric, energetic, electronic and thermodynamic properties. This is demonstrated in this study by using the new TFL representation in the molecular forward design framework as input to a (deep) convolutional neural networks (CNN) trained on $DFT_{QM}$ calculations, which outputs the properties of the molecules. Tested on the QM9 dataset (a set of 133,855 molecules and 19 properties), the new QM↔SP↔ML model is able to predict the total energies of molecules with a mean absolute error (MAE) << 1 Kcal/mol, and orbital energies with MAE << 0.1 eV, which are both below acceptable chemical accuracy. Our results also show that the new QM↔SP↔ML model performs similarly or better compared to other ML state-of-the-art techniques described in the literature. In all, in this study, we show that QM↔SP↔ML represents a powerful technique for molecular forward design.

The rest of this paper is organized as follows. Section II provides a background on QM. Section III provides a background on the forward MD using ML. Section IV describes the QM9 dataset used in this study. Section V deals with the CM and the 1D representation of molecules. Section VI presents the TFL representation of molecules. Section VII introduces the CNNs for mapping the TFL representations to molecular properties. Section VIII presents the results and discussions. This is followed by the conclusions in Section IX.

## II. Quantum Mechanics

Quantum mechanics (QM) is the science that deals with the behavior of matter and light at the atomic and subatomic scales. The Schrödinger equation (SE) is the fundamental equation of physics for describing QM systems.

$$H\Psi(r) = E\Psi(r) \tag{1}$$

where, Ψ is the state vector of the quantum system (wave function), E is the energy eigenvalue, $H = \frac{-\hbar^2}{2m}\nabla^2 + V(r)$ is the Hamiltonian, $\hbar = h/2\pi$ is the reduced Plank constant, m is the particle's mass, V(r) is the potential energy, r is the positional coordinates, and $\nabla$ is the Laplacian operator. This version corresponds to the time-independent SE. It is a partial differential equation (PDE), which uses the concept of energy conservation (Kinetic Energy + Potential Energy = Total Energy) to obtain information about the behavior of an electron bound to a nucleus. It does this by allowing an electron's wave function, Ψ, to be calculated. Solving the SE gives us Ψ and $\Psi^2$. With these, we derive the quantum numbers and the shapes and orientations of the orbitals that characterize electrons in an atom or molecule [2]. In other words, the SE account for the properties of molecules, atoms and their constituents (electrons, protons, neutrons, etc.)

Analytical or numerical solutions to the SE yield the wave function Ψ and energy E, which permit the derivation of many properties of systems. But still, many problems in materials science, organic chemistry, drug design, or biochemistry have not yet been solved. This is due to the fact that analytically, you can only solve the SE for nuclei with one electron (e.g. H, $He^{+}$, $Li^{2+}$, $Be^{3+}$, $B^{4+}$, $C^{5+}$, etc.) For all other atoms, ions, and molecules, a major problem is the computational effort required, which grows with the system size. For example, the benzene molecule ($C_6H_6$) consists of 12 nuclei and 42 electrons. The SE, which must be solved to obtain the energy and Ψ of this molecule, is a PDE in 162 variables. This situation necessitates approximate solutions in an accuracy versus generality trade-off in order to achieve computational efficiency [3]. Many such approximations were developed from both a conceptual level, such as the Born-Oppenheimer approximation, and a numerical level [4]-[13]. They lead to a multiplicity of approaches for approximately solving the SE, with different runtime [20]. $DFT_{QM}$ with a runtime of $O(N^3)$ is one of the widely used approach [10]. Here, N is the system size, e.g., number of atoms, electrons, or basis functions. To give more insights on the differences and complexities in asymptotic runtime of these methods, consider increasing a system's size N by a factor of 2. For a configuration interaction [4] and coupled cluster method with runtime $O(N^{10})$ and $O(N^7)$ [5], runtime increases by a factor of $2^{10} = 1024$ and $2^7 = 128$ respectively, whereas for a $DFT_{QM}$ [10] and molecular mechanics [12] methods with runtime $O(N^3)$ and $O(N^2)$ it increases only by a factor of 8 and 4 respectively. For large system or a large number of small systems, one might run out of computing resource using these approaches [20]. For such systems, linear-scaling QM methods offer a different approach by taking advantage of locality for an $O(N)$ asymptotic runtime [13]. But, they are not applicable to all systems [13, 20]. Another approach which is of interest in this study is to use ML for its high speed and potential for precisely ballpark QM solutions.

## III. Quantum Mechanics ↔Machine Learning Models

The ultimate goal in QM↔ML is to develop surrogate models that has the same accuracy as the SE and the high speed of ML. For example, obtaining the properties of molecules by solving the SE is computationally very expensive. As a consequence, only a small percentage of the molecules in the CCS have been labelled. By training a ML algorithm on the few labelled ones, the trained QM↔ML model can be used to predict the properties of unseen (not included in the training set) molecules. There are two types of problem in MD and ML: the forward and the inverse design. Mathematically, the forward design can be formulated as follows. Given a molecule, find its properties:

$$Properties = f\,(Molecule) \tag{2}$$

Conversely, the inverse design can be defined as follows: given the desired/targeted properties, find the molecules:

$$Molecules = f^{-1}(properties). \tag{3}$$

Our focus in this study is on the forward design problem. The inverse design problem from a SP perspective will be the subject of a subsequent paper. The function *f* in the equations above represents the relationship between the molecules and their properties, and it is inferred during the ML training step using a set of well-labelled pairs of (molecules → properties) referred to as the training set. Several ML techniques have been proposed in the literature to tackle the forward design problem.

Kernel ridge regression (KRR) [19, 20, 51], Support Vector Regression (SVR), Gaussian Process regression (GPR) [36], and Elastic Net (EN) [38, 39] have been widely used and demonstrated that, when their parameters are well-tuned they can almost reach chemical accuracy on some molecular properties. In a previous conference paper, we demonstrated without reaching chemical accuracy that the discrete Fourier transform ($DFT_{SP}$) of the 1D representation of the molecules, associated with a Gaussian KRR approach was able to produce better results compared to the 1D signal representation as input to KRR [52]. Artificial neural networks (ANN) and CNNs architectures have also been proposed and tested for the prediction of energetic and electronic properties of molecules. A Bayesian regularized NNs was shown to almost achieve chemical accuracy on the prediction of the atomization energy using the QM7 dataset [53]. A framework called Message Passing Neural Networks (MPNNs) was proposed and shown to achieve exciting performances on QM9 dataset where 11 out of 13 properties were predicted within chemical accuracy [41]. A convolutional neural networks for atomistic systems (CNNAS) was proposed for the computation of total energy of atomic systems and showed to challenge the computational cost of empirical potentials while maintaining the precision of *ab initio* results [40]. A framework that combines transferable NN potentials and a Behler-Parrinello symmetry functions called ANI was reported and showed to achieve errors in total energies prediction equal to 0.14 Kcal/mol [24]. A deep tensor NN (DTNN) to mimic many-body Hamiltonians was proposed in [42]. In the same study, the authors introduced continuous filter convolutional layers (called SchNet) as novel building blocks for deep NN [43]. The reported accuracy achieved by SchNet on QM9 is 0.32 Kcal/mol for $U_0$ and 0.04 eV and 0.03 eV for HOMO and LUMO energies respectively. A NN architecture called PhysNet was proposed in [25] and showed to reached a MAE of 0.14 Kcal/mol on total energies. The MatErials Graph Network (MEGNet), an implementation of DeepMind's graph networks [60] for universal ML in materials science was proposed in [55], and achieved very low prediction errors in a broad range of properties in both molecules and crystals. A set of computational intelligence techniques (black and white boxes) was recently tested on the QM7 dataset although they did not reach chemical accuracy, white box models brought some explainable angles to the QM↔ML problem [54].

The progress in precision achieved for energetic properties of QM9 are truly outstanding. However, much needs to be done in topics like molecular representation that captures all the features of the molecule, or in the development of new approaches for predicting a broader range of molecular properties below the acceptable chemical accuracy. Our goal in this study is to explore the MD problem from a new perspective using techniques inspired and deeply rooted into SP. The challenge is to do it within the SP framework, in a way that performs similarly or better compared to the existing state-of-the-art techniques, and also showing the advantages of using SP within the MD pipeline.

## IV. QM9 Dataset

QM9 is a comprehensive and publicly available dataset that provides geometric, energetic, electronic and thermodynamic properties for a subset of GDB-17 database, comprising 134K stable drug-like molecules that span a wide range of organic molecules. Molecules in the dataset consist of Hydrogen (H), Carbon (C), Oxygen (O), Nitrogen (N), and Fluorine (F) atoms and contain up to 9 heavy (non-Hydrogen) atoms. For each molecule $DFT_{QM}$ is used to find a reasonable low energy structure and hence atom "positions" are available.

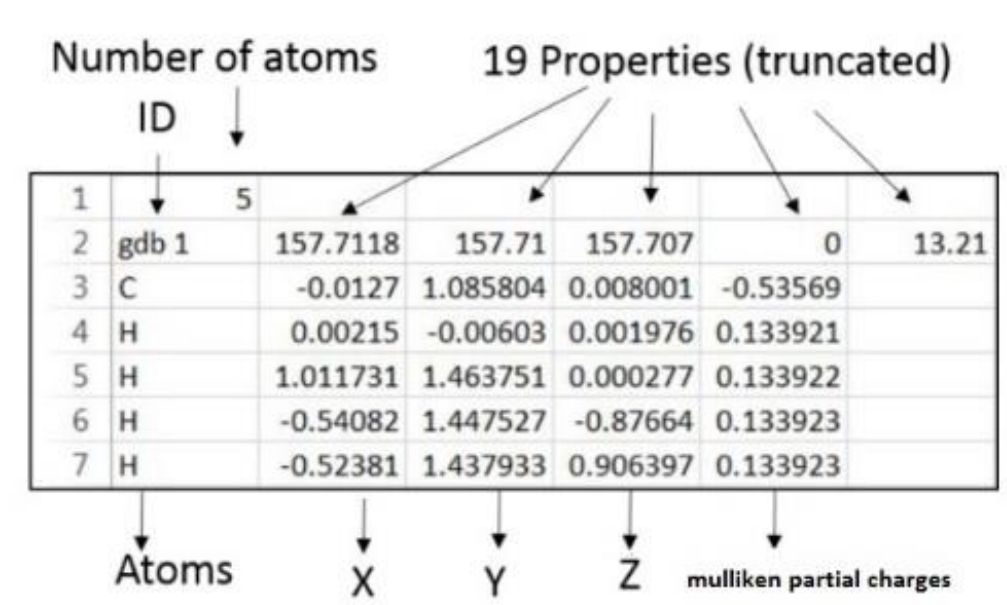

| | | | | | | |
|---|---|---|---|---|---|---|
| 1 | | 5 | | | | |
| 2 | gdb 1 | 157.7118 | 157.71 | 157.707 | 0 | 13.21 |
| 3 | C | -0.0127 | 1.085804 | 0.008001 | -0.53569 | |
| 4 | H | 0.00215 | -0.00603 | 0.001976 | 0.133921 | |
| 5 | H | 1.011731 | 1.463751 | 0.000277 | 0.133922 | |
| 6 | H | -0.54082 | 1.447527 | -0.87664 | 0.133923 | |
| 7 | H | -0.52381 | 1.437933 | 0.906397 | 0.133923 | |

**Fig. 1**. Methane $CH_4$ molecule (gdb_1) as taken from the QM9 dataset. In row 1, 5 is the number of atoms. In row 2 we have the ID of methane $CH_4$ in the database, this is followed by the properties of the molecules. Only the five first properties are shown. Then row 3 to 7, column 3 to 5 correspond to the coordinates (x, y z) of each atom while column 6 the Mulliken partial charges.

For example, **Fig. 1** shows an entry (gdb_1) of the QM9 dataset, the methane ($CH_4$) molecule. This entry describes the atomic composition of $CH_4$, its atomic coordinates and its properties computed using $DFT_{QM}$. **Fig. 2** shows a sketch of $CH_4$, with atomic number of each atom added.

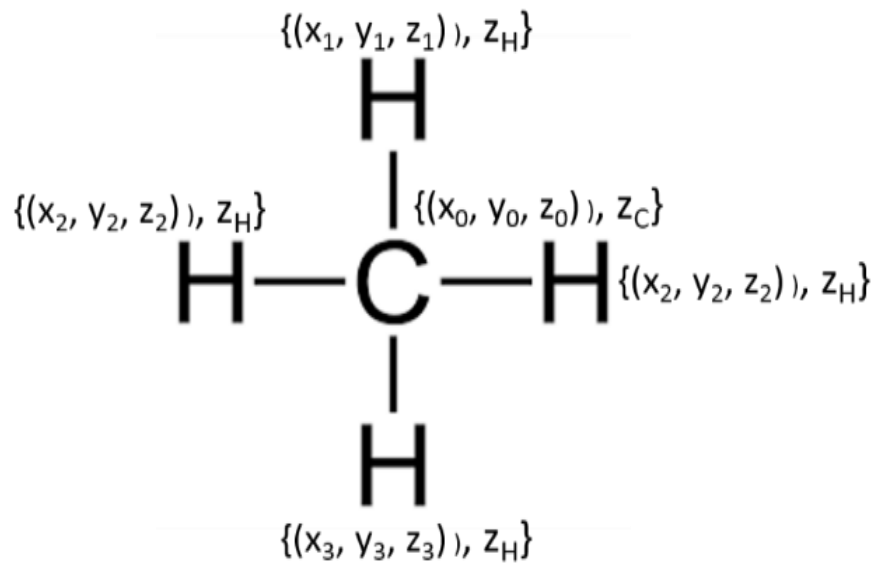

**Fig. 2**. Sketch of Methane $CH_4$ molecule (gdb 1) as taken from the QM9 dataset, the (x, y, z) represent the coordinates of the atoms, and the z the atomic number of each atom.

The version of the QM9 dataset we used has 19 properties, available in [http://moleculenet.ai/datasets-1]. We organized them in a P = [$p_{ml}$] matrix, where $p_{ml}$ is a real value that corresponds to the $l^{th}$ property of the $m^{th}$ molecule, with l = 1 to L = 19 **(Additional File 1** at: https://github.com/TABeau/QM-SP-ML). The 19 properties are: the internal energy at 0K ($U_0$), internal energy at 298.15K ($U_{298}$), Enthalpy at 298.15K ($H_{298}$), free energy at 298.15K ($G_{298}$), atomization energy at 0K ($U_0$_atom), atomization energy at 298.15K ($U_{298}$_atom), atomization enthalpy at ($H_{298}$_atom), free atomization free energy at 298.15K ($G_{298}$_atom), the zero point vibrational energy (ZPVE), the energy of the electron in the highest occupied molecular orbital (HOMO), the energy of the lowest unoccupied molecular orbital (LUMO), the electron energy gap, which is the difference HOMO – LUMO, the electronic spatial extent ($r_2$), the norm of the dipole moment (μ), the norm of static polarizability (α), the heat capacity (cv) and the rotational constants (A, B, C). For a more detailed description of these properties, see [51].

## V. Coulomb Matrix and 1D Representation of Molecules

One of the major challenges in QM↔ML is how to represent molecules in a ML pipeline. In this study, our starting point is the CM representation.

### A. *Coulomb Matrix (CM)*

Given a molecule its CM is defined by: C = [$c_{ij}$], with $c_{ij}$ defined in (4).

$$c_{ij} = \begin{cases} 0.5Z_i^{2.4} & for\ i = j \\ \frac{Z_i Z_j}{||R_i - R_j||} & for\ i \neq j \end{cases} \tag{4}$$

Where $Z_i$ is the atomic number of atom i, and $R_i$ = ($x_i$, $y_i$, $z_i$) is its position in atomic units. CM is of size I×I, where I corresponds to the number of atoms in the molecule. It is symmetric and has as many rows and columns as there are atoms in the molecule. The CM is invariant to rotation, translation but not to permutation of its atoms. Several techniques to tackle this issue have been explored in the literature. Examples include wirking with a sorted CM and with the Coulomb Eigen-spectrum (CES), which will be the one used in this study.

### B. *1D Signal of Molecules - Coulomb Eigen Spectrum (CES)*

Given C, the CM of a molecule, the CES is obtained by solving the Eigen value problem Cu = λu, under the constraints $\lambda_i > 0$, $\lambda_i \geq \lambda_{i+1}$. The spectrum ($\lambda_1, \ldots, \lambda_I$) which can be viewed as a 1D signal, is used as the representation of the molecule. Here, the 1D signal ($\lambda_1, \ldots, \lambda_I$) of the $m^{th}$ molecule ($\Omega_m$) is denoted as: x(m,:) = $x_m[n]$, with n = 1 to N. For a set of M molecules, their 1D CES signals can be organized in an M×N matrix X:

$$X = \begin{bmatrix} x_{11} & x_{12} & \ldots & x_{1n} & \ldots x_{1N} \\ x_{21} & x_{22} & \ldots & x_{2n} & \ldots x_{2N} \\ \vdots & \vdots & \vdots & \vdots & \vdots \\ x_{m1} & x_{m2} & \cdots & x_{mn} & \cdots x_{mN} \\ \vdots & \vdots & \vdots & \vdots & \vdots \\ x_{M1} & x_{M2} & \cdots & x_{Mn} & \cdots x_{MN} \end{bmatrix}. \tag{5}$$

The $m^{th}$ row of X represents the 1D signal of the $m^{th}$ molecule. Since molecules have different number of atoms, the size of the matrix will be determined by the molecule with the largest number of atoms. Accordingly, matrices corresponding to shorter molecules will be padded with zeros all of the 1D signals will then have the same length N.

## VI. Time Frequency Representation of Molecules

Time frequency representations are widely used in SP to represent, visualize and analyze signals [59]. Here, we explore these representations in the context of MD as input to a ML framework and draw hypotheses on their usefulness in molecular forward and inverse design. These transforms are referred to in this study as the time-frequency-like (TFL) transform. They do not have a time component like a typical 1D signal, but their elements form a totally ordered set (in this case the sorted eigenvalues. Note that magnitudes varying on a transect along the distance from a starting point defines 1D signals in many domains.) This study tests the short time Fourier transform, the continuous wavelet transform and the Wigner-Ville distribution.

### A. *Discrete Fourier Transform ($DFT_{SP}$)*

Given the 1D signal $x_m[n]$ of the $m^{th}$ molecule with length N, its $DFT_{SP}$ is another sequence $X_m[k]$ of the same length N (k = 0 to N-1) given by

$$X_m(k) = \sum_{n=0}^{N-1} x_m(n) e^{-j\frac{2\pi kn}{N}} \tag{6}$$

This transformation provides a measure of the frequency content at frequency k, which corresponds to an underlying period of N/k samples, where the maximum frequency corresponds to k = N/2, assuming that N is even.

### B. *Short Time Discrete Fourier Transform and Spectrogram*

The short time Fourier transform (STFT) of $x_m[n]$is obtained by applying the $DFT_{SP}$ over a sliding window w of small width to a long sequence.

$$X_{STFT}(k,l) = \sum_{n=0}^{N-1} x_m(n) w(n-k) \exp(-\frac{j2\pi nl}{N}) \quad (7)$$

$$Spectrogram(x_m(n)) = |X_{STFT}(k,l)|^2 \quad (8)$$

This equation provides a localized measure of the frequency content of $x_m[n]$. The squared magnitude of the STFT (**Eq. 8**) yields the spectrogram, which is a representation of the power spectral density of the function.

### C. *Continuous Wavelet Transform and Scalogram*

The continuous wavelet transform (CWT) of the 1D signal $x_m(t = n)$, at a scale $(a > 0)$ $a \in R^{+*}$ and translational $b \in R$ value is defined by:

$$X_{cwt}(a,b) = \frac{1}{\sqrt{|a|}} \int_{-\infty}^{+\infty} x_m(t)\, \overline{\Psi}\left(\frac{t-b}{a}\right) dt \quad (9)$$

$\Psi(t)$ is a continuous function in the time and frequency domain called the mother wavelet. The mother wavelet provides a source function that generate daughter wavelets which are simply the translated and scaled version of the mother wavelet.

$$Scalogram(x_m(t)) = |X_{cwt}(a,w)| \quad (10)$$

The scalogram is the absolute value of the CWT of $x_m[t]$, plotted as a function of time and frequency.

### D. *Wigner-Ville Distributions*

The *Wigner-Ville distribution* (WVD) provides a high-resolution time-frequency representation of a signal. For a continuous signal $x_m(t)$, the Wigner-Ville distribution is defined as:

$$WVD_{x_m}(t,f) = \int_{-\infty}^{+\infty} x_m(t+\frac{\tau}{2}) x_m^*(t-\frac{\tau}{2}) e^{-j2\pi f\tau} d\tau \quad (11)$$

For a discrete signal with $N$ samples, the distribution becomes

$$WVD_{x_m}(n,k) = \sum_{q=-N}^{N} x_m(n+\frac{q}{2}) x_m^*(n-\frac{q}{2}) e^{-j2\pi kq/N.} \quad (12)$$

From **Eq. 11** and **12**, one can notice that the WVD computes the Fourier transform of the autocorrelation function.

## VII. Learning the Mapping Between Time Frequency Representation and Properties of Molecules: (Deep) Convolutional Neural Networks

In the solution of the direct problem, molecular structures are first converted to their CMs, next to their CESs, and are finally modeled using the TFL representations as defined above. The TFLs correspond to the input of the system (CNNs in this case), while the properties correspond to its output, **Fig**. 3. The objective is to learn a mapping between the TFL representations (2D images) of a molecule and their properties (scalar). From a mathematical and ML perspective, this is a regression problem and it is tackled here using (deep) CNNs. Deep CNNs are computational architectures introduced in [61]. They have been shown to provide extraordinary regression and classification results in high dimension [62]-[63]. There is a huge literature relative to (deep) CNNs. A good description of these computational architectures can be found in [64].

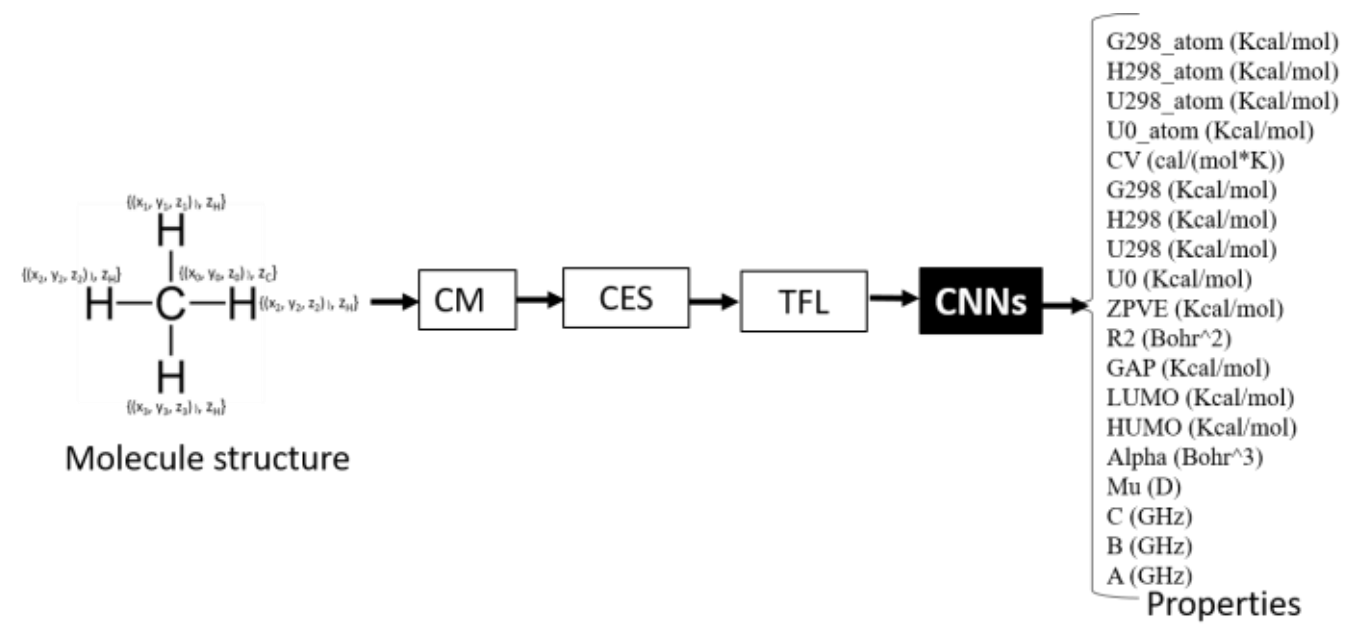

**Fig. 3**. Illustration of the QM↔SP↔ML framework using methane cartoon representation.

## VIII. Results and Discussions

The CES of each molecule was computed using their atomic coordinates as described in the QM9 dataset and the approach described above. They were then organized in an M×N = 133885×29 matrix (**Additional File 2** at: https://github.com/TABeau/QM-SP-ML). M = 133885 corresponds to the number of molecules in the QM9 dataset and N = 29 the number of atoms in the largest molecule. As mentioned in Section V, molecules with less than 29 atoms were padded with zeros so that all the 1D signals have the same dimension (N = 29). The STFT used a Hamming window, the CWT a Morlet (Gabor) wavelet, and the WVD of each molecule was computed using the Matlab script provided as **Additional File 3** at: https://github.com/TABeau/QM-SP-ML.

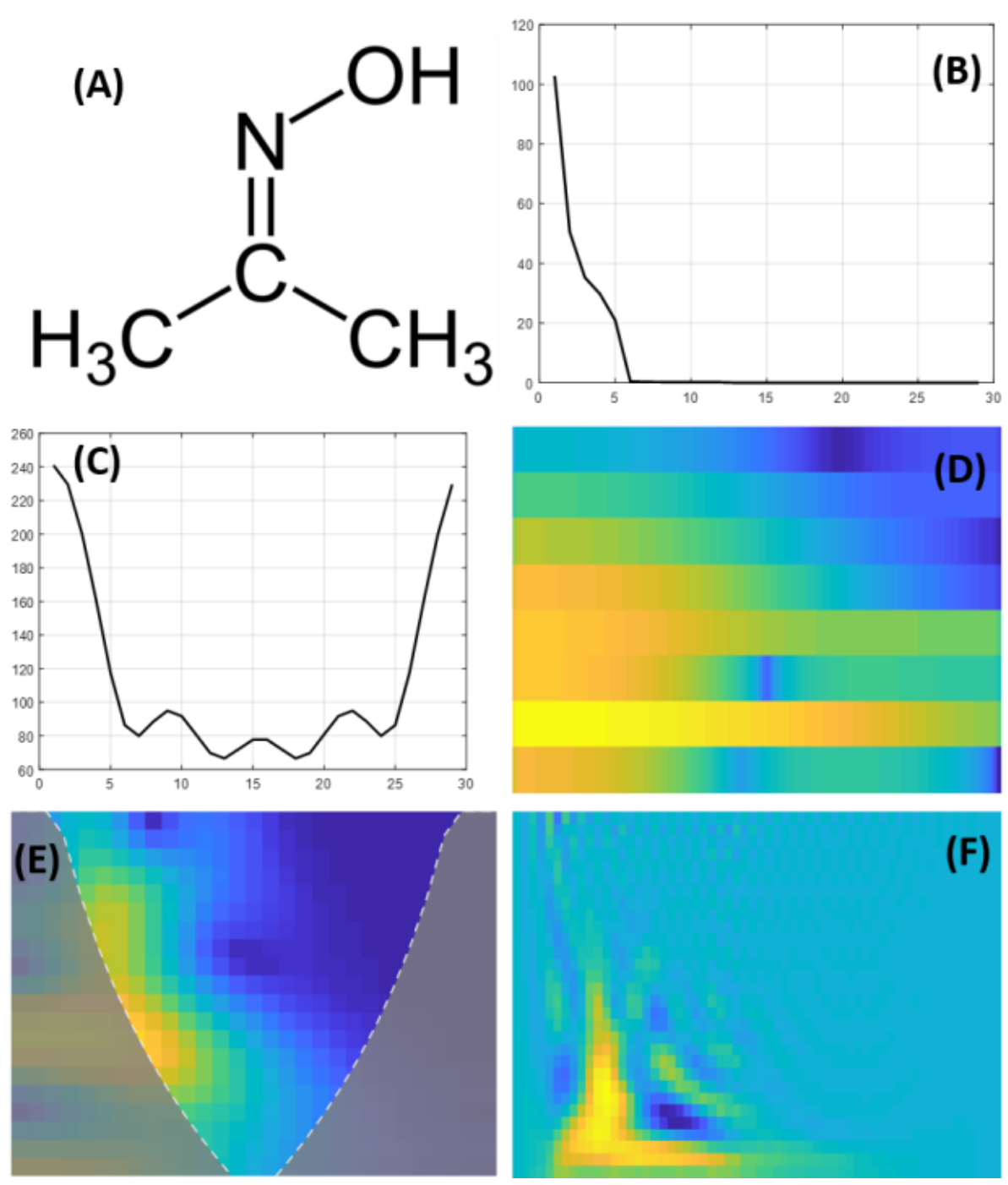

**Fig. 4.** (**A**) Chemical representation of molecule ID gdb_49 in the QM9 dataset which corresponds to one of the isomers of $C_3H_7NO$, (**B**) its 1D signal, (**C**) the amplitude of its discrete Fourier transform, (**D**) its Spectrogram (amplitude of its STFT), (E) its Scalogram (CWT) and (D) its Wigner-Ville Distribution (WVD).

As an example, **Fig. 4** illustrates the case of molecule $C_3H_7NO$ (ID = gdb_49 in the QM9 dataset). (A) is the molecule, (B) its 1D signal according to the aforementioned representation procedure, (C) the amplitude of its 1D $DFT_{SP}$, (D) its Spectrogram, (E) its Scalogram and (F) its WVD, respectively.

The dataset was randomly divided into 90% (120 500 ≈ 120K) for training and the remaining 10% (13 389 ≈ 13K) for testing. A deep CNNs was constructed using the Python script provided as **Additional File 4** at: https://github.com/TABeau/QM-SP-ML. Readers can refer to this file for details relative to the construction of the deep CNNs. Training of each TFL representation was performed on three different machines with GPU (NVIDIA Quadro K2200, NVIDIA Quadro P2000, NVIDIA GeForce GTX TITAN X) capabilities and took 3, 2 to 1 weeks for completion respectively. Performance of the $n^{th}$ property is measured using the mean absolute error (MAE)

$$MAE_n = \frac{1}{M}\sum_{m=1}^{M} |P_{mn} - P_{mn}^{e}| \ . \qquad (16)$$

$P_{mn}$ is the measured $n^{th}$ property of the $m^{th}$ molecule, and $P_{mn}^{e}$ the estimated one.

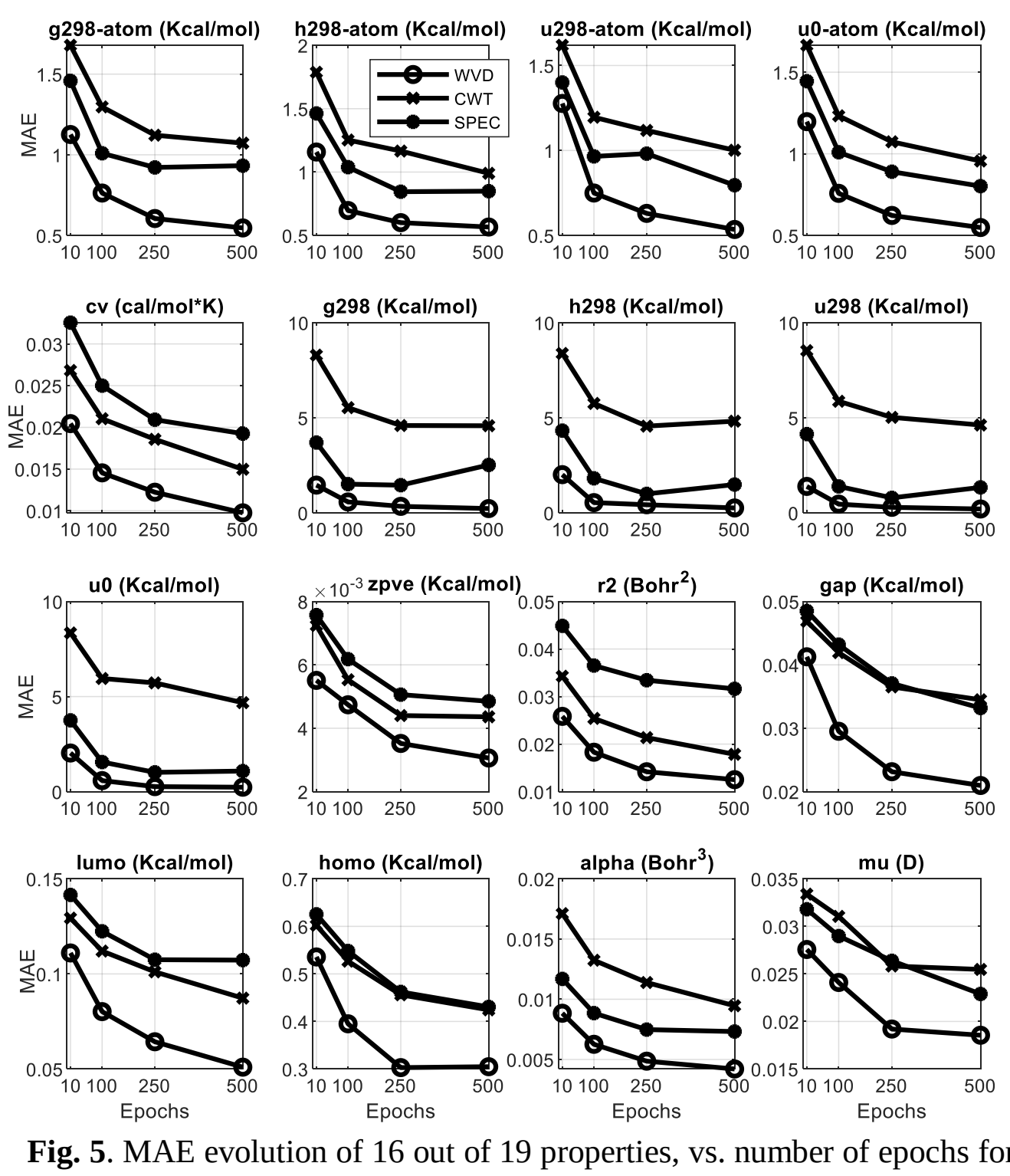

**Fig. 5**. MAE evolution of 16 out of 19 properties, vs. number of epochs for each time-frequency like representation during the training stage. The Y-axis correspond to the MAEs and the X-axis to the number of epochs.

**Fig. 5** and **Fig. 6** show the training and testing results obtained for 10, 100, 250, and 500 epochs for 16 out of the 19 properties, for WVD, CWT and STFT respectively. The best results for each TFL representation (i.e. the MAE obtained prior to the model starts overfitting) are presented in **Table III**. It is interesting to note that several of these properties are predicted with MAE below chemical accuracy.

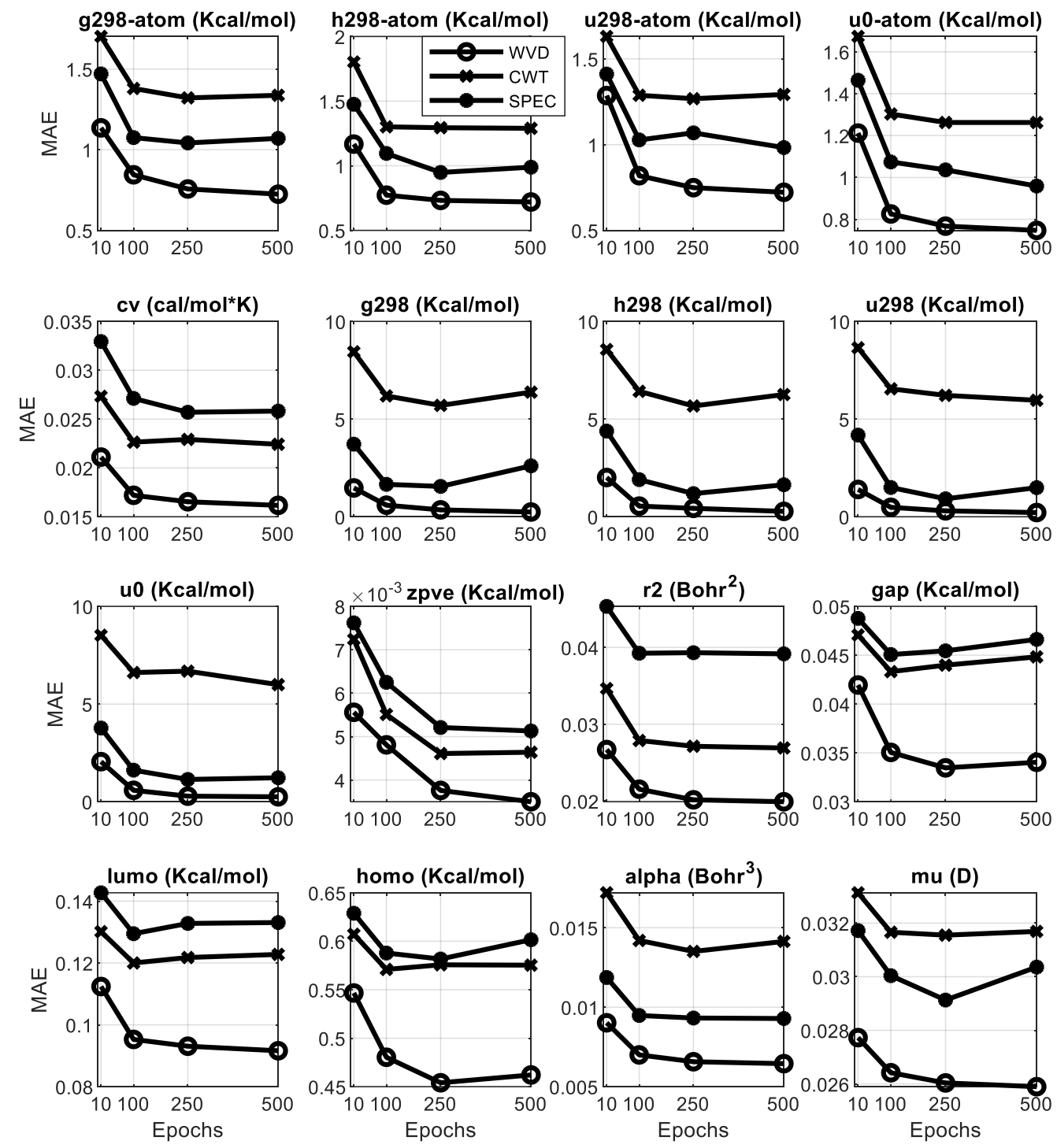

**Fig. 6**. MAE evolution of the 16 out of 19 properties, with number of epochs for each time-frequency-like representations during the testing stage. The Y-axis correspond to the MAEs and the X-axis to the number of epochs.

**TABLE III**

MEAN ABSOLUTE ERRORS OF THE THREE REPRESENTATIONS ON THE TESTING SET.

| | | MAE | | | Epochs |
|---|---|---|---|---|---|
| Properties | Unit | STFT | CWT | WVD | [STFT CWT WVD] |
| g298_atom | kcal/mol | 1.042 | 1.321 | 0.724 | [250 250 500] |
| h298_atom | kcal/mol | 0.948 | 1.294 | 0.719 | [250 250 500] |
| u298_atom | kcal/mol | 0.982 | 1.292 | 0.722 | [500 250 500] |
| u0_atom | kcal/mol | 0.958 | 1.262 | 0.747 | [500 500 500] |
| cv | cal/(mol*K) | 0.025 | 0.022 | 0.016 | [500 500 500] |
| g298 | kcal/mol | 1.554 | 5.701 | 0.244 | [250 250 500] |
| h298 | kcal/mol | 1.186 | 5.671 | 0.277 | [250 250 500] |
| u298 | kcal/mol | 0.921 | 6.214 | 0.216 | [250 500 500] |
| u0 | kcal/mol | 1.141 | 6.684 | 0.251 | [250 500 500] |
| zpve | kcal/mol | 0.005 | 0.004 | 0.003 | [500 250 500] |
| r2 | Bohr^2 | 0.039 | 0.026 | 0.019 | [500 500 500] |
| gap | kcal/mol | 0.045 | 0.043 | 0.033 | [100 100 250] |
| lumo | kcal/mol | 0.129 | 0.120 | 0.091 | [100 100 500] |
| homo | kcal/mol | 0.581 | 0.570 | 0.454 | [250 100 250] |
| alpha | Bohr^3 | 0.009 | 0.013 | 0.006 | [500 250 500] |
| mu | D | 0.029 | 0.031 | 0.025 | [250 250 500] |
| C | GHz | - | - | - | - |
| B | GHz | - | - | - | - |
| A | GHz | - | - | - | - |

The epochs column = [STFT CWT WVD] specifies the number of epochs where each representation achieved the best MAE result respectively, prior to the model starts overfitting.

**Fig. 7**, **Fig. 8** and **Fig. 9** shows the combined MAE evolution of training and testing for STFT, CWT and WVD on the same graph respectively. These figures show a better description of when the corresponding model starts overfitting. For example, for the LUMO property, the model corresponding to the STFT and CWT representations start to overfit after 100 epochs, whereas the one corresponding to WVD keeps improving up to 500 epochs.

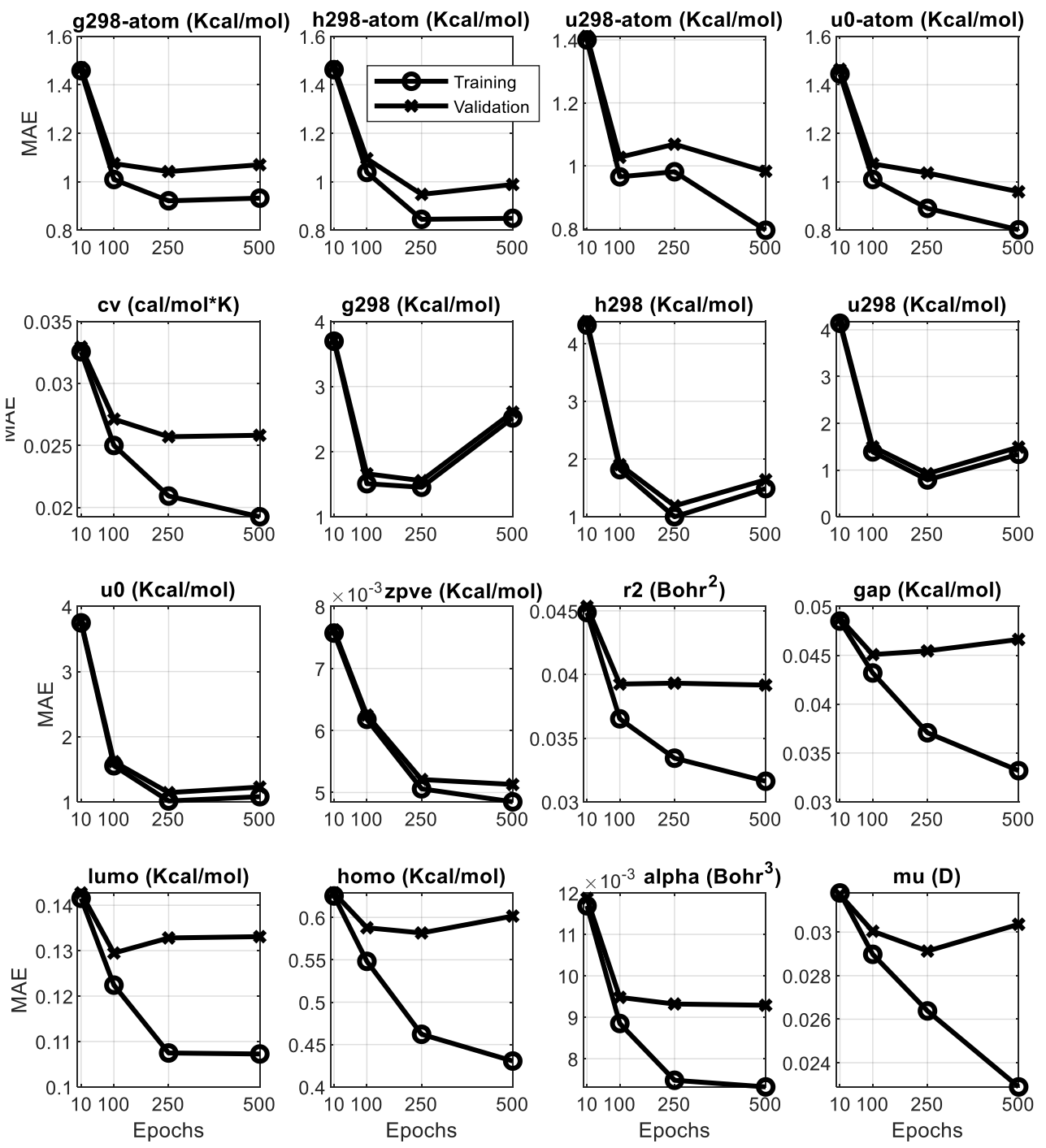

**Fig. 7**. MAE evolution of the 16 out of 19 properties, with number of epochs for the STFT/Spectrogram during the training and testing stage. The Y-axis correspond to the MAEs and the X-axis to the number of epochs.

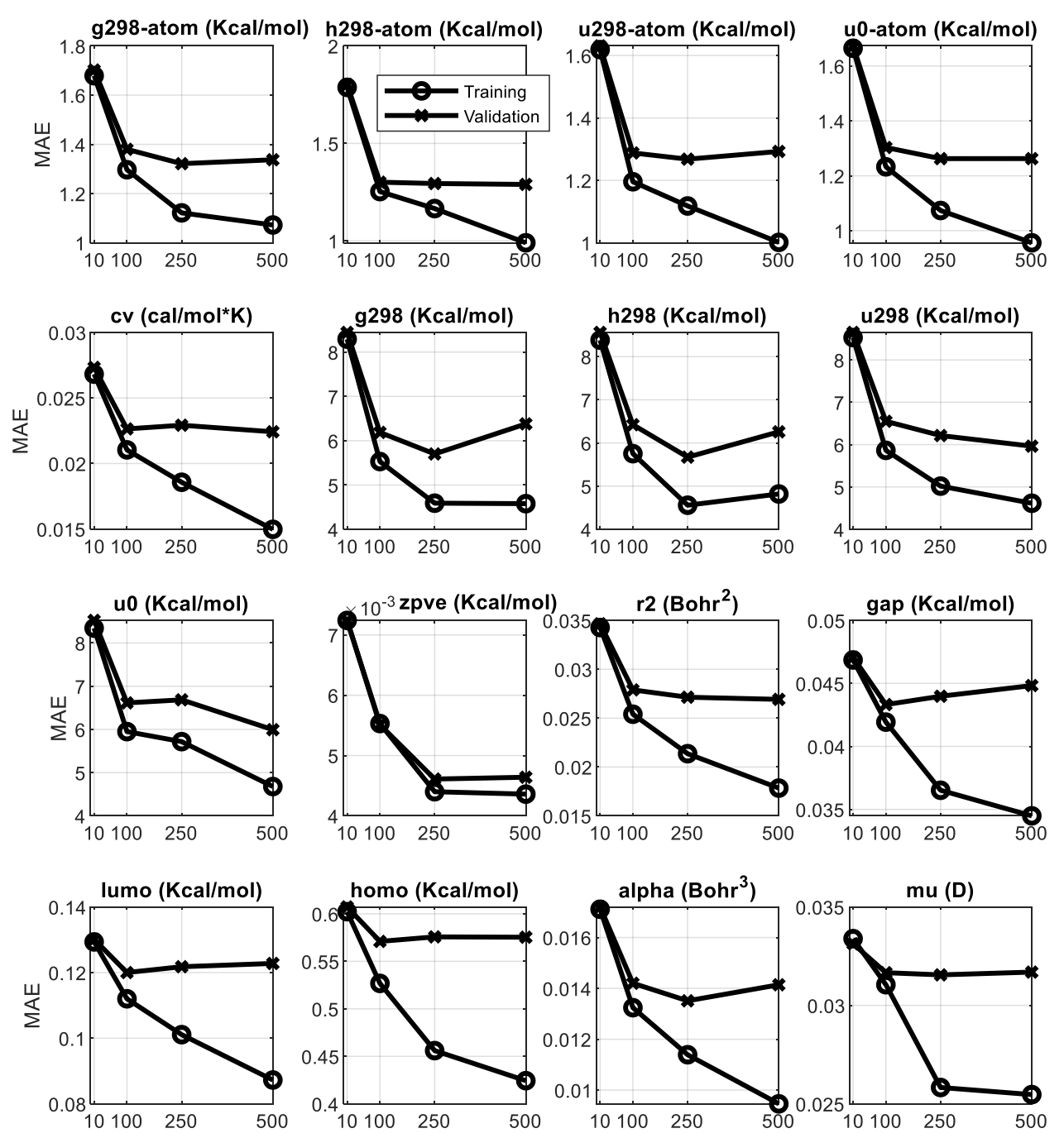

**Fig. 8**. MAE evolution of the 16 out of 19 properties, with number of epochs for the Scalogram/continuous wavelet transform during the training and testing stage. The Y-axis correspond to the MAEs and the X-axis to the number of epochs.

By running the training above 500 Epochs, the accuracy of some of these properties can be further improved.

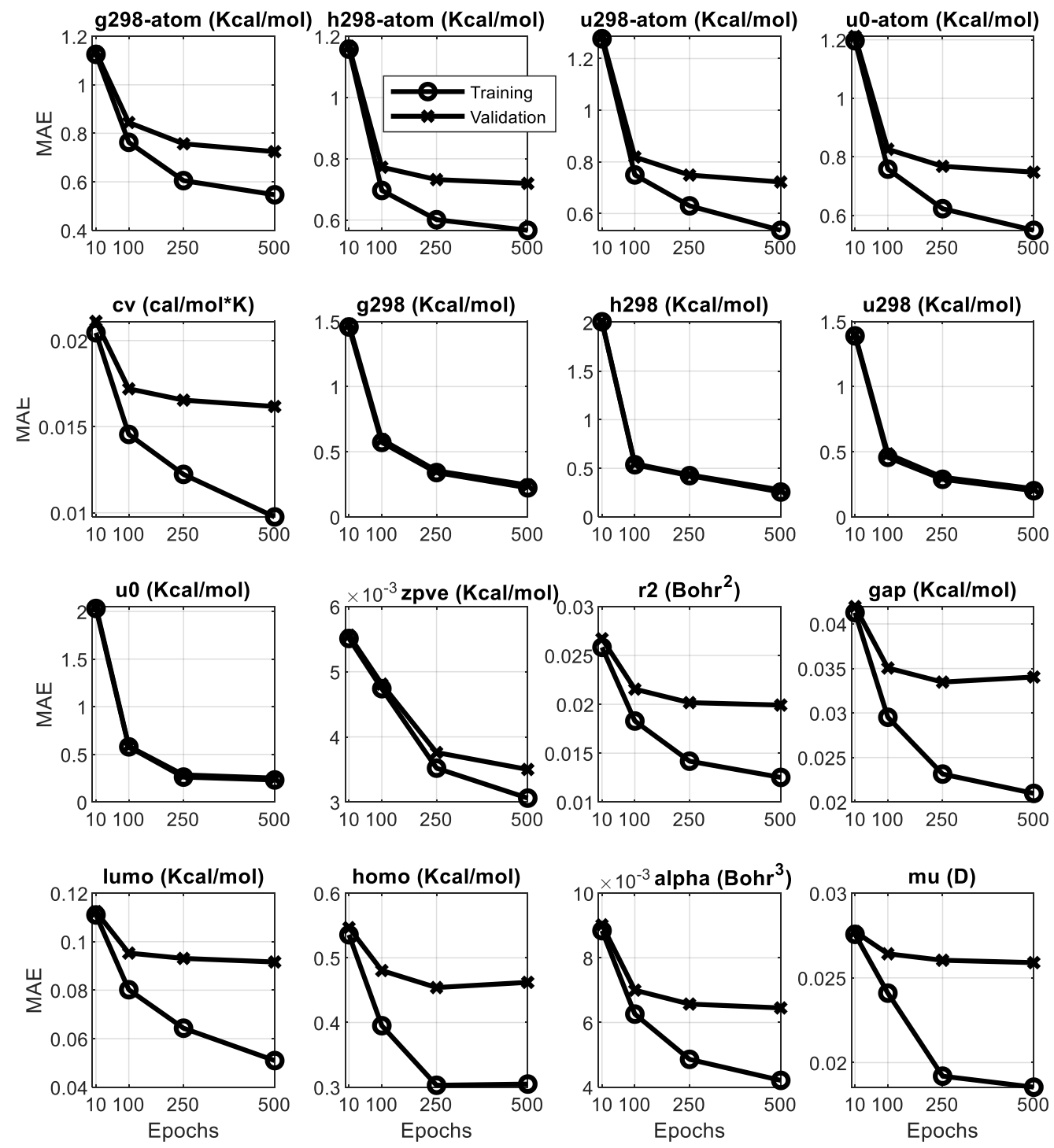

**Fig. 9**. MAE evolution of the 16 out of 19 properties, with number of epochs for the Wigner-Ville Distribution (WVD) during the training and testing stage. The Y-axis correspond to the MAEs and the X-axis to the number of epochs.

### *A. Comparison between STFT, CWT, and WVD*

Among the three representations, the model relative to the WVD gave the best training and testing set prediction results for all the 19 properties and for models at 10, 100, 250 and 500 epochs compared to the STFT and CWT. The STFT came second and the CWT third. More precisely, the WVD predicted 16 properties out of 19 with MAEs bellow chemical accuracy. The STFT predicted 16 out of 19 with 12 MAEs below chemical accuracy and 4 equal or slightly above chemical accuracy. The CWT performed the worst. It predicted 16 out of 19 properties with only 8 properties below chemical accuracy.

### *B. Comparison between QM↔SP↔ML and other ML Techniques*

**Table IV** gives a comparative analysis of the QM↔SP↔ML method and the state-of-the-art ML techniques described in the literature and mentioned in **Table II** above. On the $G_{298}$_atom, $H_{298}$_atom, $U_{298}$_atom and $U_0$_atom, the WVD scored a MAE of around 0.7 Kcal/mol, which is < 1 Kcal/mol. There were no other available ML results in the literature for comparison. On the $G_{298}$, $H_{298}$, and $U_{298}$, our approach via the WVD was slightly better compared to the results mentioned in the literature. We obtained MAEs of 0.244Kcal/mol, 0.277Kcal/mol, 0.216Kcal/mol compared to 0.276 Kcal/mol, 0.276Kcal/mol and 0.299Kcal/mol of the MEGNet algorithm respectively. On the $U_0$, among the six ML approaches that we compared the QM↔SP↔ML to, the WVD came second with a MAE of 0.25 Kcal/mol slightly higher than the 0.14Kcal/mol obtained by the SOAP algorithm [49] and the PhysNet algorithm [25].

**TABLE IV**
MAE VALUES ON THE TESTING SET OF THE COMPARATIVE ANALYSIS OF THE QM↔SP↔ML APPROACH WITH STATE-OF-THE-ART ML TECHNIQUES FOR THE 19 PROPERTIES OF THE QM9 DATASET.

| Properties | Units | QM-SP-ML | MEGNet | KRR/ BAML | GPR/SOAP/ GAP | NMP | Multitask | SchNet | HIP-NN | HDNN | KRR/SOAP | PhysNet |
|---|---|---|---|---|---|---|---|---|---|---|---|---|
| g298_atom | kcal/mol | **0.724** | - | - | - | - | - | - | - | - | - | - |
| h298_atom | kcal/mol | **0.719** | - | - | - | - | - | - | - | - | - | - |
| u298_atom | kcal/mol | **0.722** | - | - | - | - | - | - | - | - | - | - |
| u0_atom | kcal/mol | **0.747** | - | - | - | - | - | - | - | - | - | - |
| cv | cal/(mol*K) | **0.016** | 0.029 | 1.64 | - | 0.80 | 0.124 | 0.033 | - | - | - | - |
| g298 | kcal/mol | **0.244** | 0.276 | 1.20 | - | 0.44 | 44.32 | 0.322 | - | - | - | - |
| h298 | kcal/mol | **0.277** | **0.276** | 1.22 | - | 0.39 | 44.16 | 0.322 | - | - | - | - |
| u298 | kcal/mol | **0.216** | 0.299 | 1.22 | - | 0.45 | 43.96 | 0.438 | - | - | - | - |
| u0 | kcal/mol | 0.251 | 0.276 | 1.21 | 0.28 | 0.45 | 44.04 | 0.32 | 0.26 | 0.41 | **0.14** | **0.14** |
| zpve | kcal/mol | 3e-3 | **3e-5** | 3.31 | - | 1.27 | 0.199 | **3e-5** | - | - | - | - |
| r2 | Bohr^2 | **0.019** | 0.302 | 3.25 | - | 0.15 | 2.056 | 0.073 | - | - | - | - |
| gap | kcal/mol | **0.033** | 1.522 | 3.28 | - | 1.60 | 2.014 | 1.452 | - | - | - | - |
| lumo | kcal/mol | **0.091** | 1.014 | 2.76 | - | 0.87 | 1.133 | 0.691 | - | - | - | - |
| homo | kcal/mol | **0.454** | 0.991 | 2.20 | - | 0.99 | 1.620 | 0.922 | - | - | - | - |
| alpha | Bohr^3 | **0.006** | 0.081 | 3.01 | - | 0.92 | 0.571 | 0.235 | - | - | - | - |
| mu | D | **0.025** | 0.050 | 4.34 | - | 0.30 | 0.304 | 0.033 | - | - | - | - |
| C | GHz | - | - | - | - | - | 0.009 | - | - | - | - | - |
| B | GHz | - | - | - | - | - | 0.016 | - | - | - | - | - |
| A | GHz | - | - | - | - | - | 0.099 | - | - | - | - | - |

On the zpve, our approach score a MAE of 3e-3Kcal/mol and came third compared to the 3e-5Kcal/mol of MEGNet and SchNet. On cv, $r_2$, gap, LUMO, HOMO, alpha and mu properties, our three representations (STFT, CWT, and WVD) gave better results compared to the ones mentioned in the literature. Finally, on the C (rotational constant), B (rotational constant) and A (rotational constant) our methods failed to predict compared to the MAEs of 0.009, 0.016, 0.099 GHz obtained by the multitask NN algorithm [44].

It is interesting to outline the superiority of the QM↔SP↔ML model on the prediction of properties such as: r2, gap, LUMO, HOMO alpha and mu. In the case of the gap property for example, the QM↔SP↔ML model score a MAE = 0.033 kcal/mol, with the SchNet algorithm coming second with a MAE = 1.452kcal/mol. That is an order of magnitude 1.451/0.33 = 44 higher than that of the QM↔SP↔ML model. Similar conclusion can be drawn for r2, LUMO, HOMO alpha and mu. In all the new proposed QM↔SP↔ML model via the WVD representation outperforms several of the state-of-the-art ML techniques described in the literature on the prediction of 14 properties and was able to predict 16 out of 19 properties of the QM9 dataset with MAEs below chemical accuracy.

### *C. What Information are Encoded in the Time-Frequency Representations?*

The success of the TFL representations of molecules in the prediction of their properties with MAEs below chemical accuracy mean that these representations encode very relevant information pertaining to the molecules. The connection between the TFL representations and the structure of the molecule is obvious because the TFL representations are inferred from the CM which are computed using the atomic coordinates. Note that the CM is directly derived from the geometry representation of molecules. It is well known that the structure of a molecule dictates its properties. This structure to property relationship combined with the fact that the TFL representations are able to predict the properties of molecules with MAEs below chemical accuracy further validate the assertion that chemical knowledge is indeed encoded in them. Another question that might come up is, why not just use the 1D signal representation (i.e. CES) and not the TFL representation as input to ML framework? Why taking this extra step to convert the 1D numerical signal to a 2D image signal? The Multitask NN algorithm [44] did just that. In the multitask NN the 1D CES representation of molecule is used as input to a deep multitask NN. As we showed in this study (**Table IV**), the new QM↔SP↔ML model based on image representation outperformed the multitask NN on 16 properties out of 19. For example our algorithm predicted $G_{298}$, $H_{298}$, $U_{298}$, and $U_0$ with MAEs below chemical accuracy whereas the multitask NN scored ~ 44Kcal/mol, way above chemical accuracy. This is a very big difference and further validate the extra step of converting the 1D signal into a 2D representation. The fact that the TFL representations perform better than the 1D CES suggests that information that were not obvious in the 1D signal are amplified and made explicit in the 2D image representations. In audio SP for example, it is well acknowledged that the appearance of spectrograms encloses significant information about signals, to the point that experts can infer the words uttered in audio signals by simple visual examination of their spectrograms [59].

## IX. CONCLUSIONS

In this study, we showed that time-frequency-*like* representations of molecules is a powerful tool that can be used for molecular representation and visualization. We demonstrated that these representations encode the structural, geometric, energetic, electronic and thermodynamic properties of molecules. Using a deep convolutional neural networks approach in a regression framework and the benchmark QM9 dataset, we showed that there exist a clear relationship between the time-frequency-*like* representations and the structure, energetic, electronic, and thermodynamic properties of the molecules. All the codes and data generated and used in this study are available as supporting documents. **Additional File 5** at: https://github.com/TABeau/QM-SP-ML contains the Molecules ID. The readme file contains a detail description of all the additional files and how to set the Matlab codes, Python scripts, and different files and folders to run on a computer.

## ACKNOWLEDGMENT

This work is supported by the National Research Council of Canada through its Artificial Intelligence for Design Program led by the Digital Technologies Research Centre.